\documentclass[11pt]{article}

\usepackage{graphicx,enumitem}
\usepackage{graphicx,psfrag,epsf}
{%
\centering\addtocounter{figure}{1}
\begin{enumerate}[%
itemsep=2pt,parsep=0em,
label={(\alph*)},
ref={\thefigure.(\alph*)}
]}%
{\end{enumerate}\addtocounter{figure}{-1}}

\usepackage[margin=1in]{geometry}

\usepackage[utf8]{inputenc}
\usepackage{amsmath}
\usepackage[english]{babel}
\usepackage{amssymb}
\usepackage{amsfonts}
\usepackage{bm}
\usepackage{graphicx}
\usepackage{amsmath}
\usepackage{amsthm}
\usepackage{xcolor}
\usepackage[hidelinks=true]{hyperref}
\usepackage{parskip}
\usepackage{multicol}
\usepackage{float}

\usepackage{verbatim}

\floatstyle{plaintop}
\restylefloat{table}
\usepackage[tableposition=top]{caption}

\usepackage{rotating}
\usepackage{subcaption}
\usepackage{multirow}

\usepackage{chngcntr}
\usepackage{apptools}
\AtAppendix{\counterwithin{lem}{subsection}}
\AtAppendix{\counterwithin{proposition}{subsection}}
\AtAppendix{\counterwithin{equation}{subsection}}

 \theoremstyle{definition}

\newtheorem{proposition}{Proposition}[section]
\newtheorem{mydef}{Definition}[section]
\newtheorem*{prueba}{Proof}

\usepackage{booktabs}
\usepackage{animate}

\usepackage{natbib}
\hypersetup{urlcolor=blue, citecolor=red, colorlinks=true, linkcolor=red} 

\numberwithin{equation}{section}
\numberwithin{figure}{section}
\numberwithin{table}{section}

\renewcommand{\qed}{\hfill\blacksquare}

\usepackage{authblk}

\usepackage{blindtext}
\usepackage{lineno}

\begin{document}

\def\spacingset#1{\renewcommand{\baselinestretch}%
{#1}\small\normalsize} \spacingset{1}


\title{Effective Sample Size for Functional Spatial Data}

\author{
Alfredo Alegr\'ia$^{1}$\thanks{Corresponding author. Email: alfredo.alegria@uc.cl}, 
John G\'omez$^{2}$, 
Jorge Mateu$^{3}$, and 
Ronny Vallejos$^{2}$
}

\date{}

\maketitle

\vspace{-1.5em}
\begin{center}
{\small
$^{1}$Departamento de Estad\'istica, Pontificia Universidad Cat\'olica de Chile, Santiago, Chile\\
$^{2}$Departamento de Matem\'atica, Universidad T\'ecnica Federico Santa Mar\'ia, Valpara\'iso, Chile\\ 
$^{3}$Department of Mathematics, University Jaume I, Castell\'on, Spain
}

\vspace{1em}
{\small\itshape \today}
\vspace{1em}
\end{center}

\begin{abstract}
\noindent The effective sample size quantifies the amount of independent information contained in a dataset, accounting for redundancy due to correlation between observations. While widely used in geostatistics for scalar data, its extension to functional spatial data has remained largely unexplored. In this work, we introduce a novel definition of the effective sample size for functional geostatistical data, employing the trace-covariogram as a measure of correlation, and show that it retains the intuitive properties of the classical scalar ESS. We illustrate the behavior of this measure using a functional autoregressive process, demonstrating how serial dependence and the allocation of variability across eigen-directions influence the resulting functional ESS. Finally, the approach is applied to a real meteorological dataset of geometric vertical velocities over a portion of the Earth, showing how the method can quantify redundancy and determine the effective number of independent curves in functional spatial datasets. \\

\end{abstract}

\section{Introduction}

The concept of Effective Sample Size (ESS) provides a way to quantify the amount of independent information contained in a dataset. Since its early development, various forms of ESS have been introduced across different areas of statistics. In stochastic simulation, it plays a central role in evaluating the performance of algorithms such as Markov chain Monte Carlo and importance sampling \citep{gamerman2006markov,martino2017effective,chatterjee2018sample}. In model selection, it contributes to reliable assessments of model complexity \citep{Berger:2014}, while in time series analysis, it allows standard hypothesis test statistics to be properly adjusted to account for autocorrelation \citep{yue2004mann}. 

A particularly relevant setting for ESS is geostatistics, where spatial autocorrelation often causes the nominal sample size to exaggerate the true amount of information in the data \citep{cressie2015statistics}. In this context, ESS provides a means to adjust for spatial redundancy, leading to efficient subsampling, and reductions in computational burden.   Practical applications span topics such as soil contamination studies \citep{Griffith:2005}, vegetation analysis \citep{Li:2016}, and biodiversity research based on forest imagery \citep{vallejos2014effective,acosta2021assessing}. Extensions of ESS methodology to spatial regression models \citep{acosta2018effective2}, space--time data \citep{alegria2023effective} and multivariate spatial data \citep{vallejos2021effective} further demonstrate its broad applicability.

In parallel, Functional Data Analysis (FDA) has developed as a major area of statistics concerned with data that take the form of  functions \citep{ramsay2008functional}. Each functional observation represents the realization of an underlying random process, and statistical methods must therefore account for the particular nature of the data. Many datasets consist of functions observed over a continuum, and treating them as multivariate vectors ignores their smoothness and the intrinsic dependence structure induced by the continuum. FDA addresses this challenge by representing each function with a smooth basis, reducing noise and dimensionality while preserving the underlying functional structure.  Applications of FDA are widespread, encompassing medicine, climatology, and environmental science, among others \citep{febrero2008outlier,suhaila2017spatial,orozco2025functional}.

These two areas, spatial statistics and FDA, intersect in the study of functional spatial data, characterized by functions indexed by spatial location \citep{delicado2010statistics, mateu2022introduction}. Such data arise naturally in environmental monitoring, precision agriculture, and neuroscience, where processes evolve continuously over one dimension, such as time or vertical position, and exhibit spatial dependence. A growing literature has addressed data visualization \citep{sun2012adjusted}, modeling, inference, and spatial interpolation \citep{giraldo2010continuous,giraldo2011ordinary,caballero2013universal,menafoglio2013universal,sguera2014spatial,bernardi2017penalized,hernandez2020recent,yarger2022functional}, as well as spectral theory in this setting \citep{caponera2024asymptotics,kartsioukas2023spectral}.

Although effective sample size methodology and functional spatial statistics have advanced, a formal definition of ESS for functional spatial data has not yet been established. Developing a functional ESS requires accounting for both the functional nature of the observations and their spatial dependence. We adopt a Hilbert-space framework, representing each observation as a function in the space $L^2([0,1])$ of square-integrable functions over the interval $[0,1]$. Spatial dependence is described by a covariance operator, and the trace-covariogram provides a practical scalar summary of this operator. Building on this framework, we define a functional ESS that preserves the key properties and interpretability of the scalar case: it quantifies the number of independent function-valued observations that would provide the same information as the current sample. We also describe practical methods for its estimation. To illustrate the behavior of this measure, we first consider a simple functional autoregressive process as a concrete example. Finally, we apply our approach to a real dataset of geometric vertical velocities observed over a portion of the Pacific Ocean, demonstrating its practical relevance.

The remainder of the paper is organized as follows. Section \ref{sec:background} provides background on ESS for scalar random fields and on functional spatial data. Section \ref{sec:results} introduces the functional  ESS and presents its theoretical properties, estimation procedures, and illustrative examples. Section \ref{sec:data} demonstrates the methodology with a real data application, and Section \ref{sec:conclusions} concludes with a discussion and directions for future research.

\section{Background}
\label{sec:background}

\subsection{ESS for scalar random fields}

Let \(\{ X_{{\bm{s}}} : {\bm{s}} \in \mathbb{R}^d \}\) denote a real-valued random field on $\mathbb{R}^d$, with $d\in\mathbb{N}$. We focus on second-order stationary random fields, meaning that each random variable \(X_{{\bm{s}}}\) has a finite variance, that there exists a constant \(\mu \in \mathbb{R}\) such that \(E[X_{{\bm{s}}}] = \mu\) for all \({\bm{s}} \in \mathbb{R}^d\), and that, for any pair of spatial locations \({\bm{s}}, {\bm{s}}' \in \mathbb{R}^d\), the covariance function takes the form
\[
\text{cov}\big[ X_{{\bm{s}}}, X_{{\bm{s}}'} \big] = \sigma^2 r(\bm{h}; \bm{\alpha}), \qquad \bm{h} = {\bm{s}} - {\bm{s}}' \in \mathbb{R}^d,
\]
where \(r(\cdot; \bm{\alpha}) : \mathbb{R}^d \rightarrow [-1,1]\) is a parametric correlation function depending only on the spatial separation \(\bm{h}\) and on a parameter vector \(\bm{\alpha}\) \citep{chiles2009geostatistics}. Here, \(\sigma^2 > 0\) regulates the variance of the random field. If the covariance depends on \(\bm{h}\) only through its Euclidean norm \(h=\|\bm{h}\|\), the random field is termed isotropic. In this case, $\bm{\alpha}$ typically contains parameters governing the rate of correlation decay and the shape of the covariance function, while a nugget effect may be added to account for measurement error.

Suppose that the random field is observed at a set of locations \(\{{\bm{s}}_1, \dots, {\bm{s}}_n\} \subset \mathbb{R}^d\). Let \({\bf R}(\bm{\alpha})\) denote the \(n \times n\) correlation matrix of the random vector \(\bm{X} = [X_{{\bm{s}}_1}, \dots, X_{{\bm{s}}_n}]^\top\), where $\top$ denotes transposition. Then,
\[
{\bf R}(\bm{\alpha}) = \left[ r({\bm{s}}_i - {\bm{s}}_j; \bm{\alpha}) \right]_{i,j=1}^n,
\]
which is symmetric and positive semidefinite. The ESS quantifies the effective number of independent observations in \(\bm{X}\). A widely used definition, due to \cite{Griffith:2005}, is based on a variance inflation factor comparing the variance of the sample mean under independence with that under spatial correlation.

\begin{mydef}[\citealp{Griffith:2005}]
\label{def_ess}
Let \({\bf R}(\bm{\alpha})\) be the correlation matrix of \(\bm{X} = [X_{{\bm{s}}_1}, \dots, X_{{\bm{s}}_n}]^\top\), and let \(\bm{1}_n = [1, \dots, 1]^\top\). The ESS is defined as
\begin{equation}
    \label{ess} 
    \text{ESS} = n^2 \left( \bm{1}_n^\top {\bf R}(\bm{\alpha}) \bm{1}_n \right)^{-1} =  n^2 \left(\sum_{i=1}^n \sum_{j=1}^n r({\bm{s}}_i - {\bm{s}}_j; \bm{\alpha}) \right)^{-1}.
\end{equation}
\end{mydef}

Alternative definitions have been proposed using Fisher information \citep{vallejos2014effective} and Godambe information \citep{acosta2021assessing}. While these variants generally lead to similar behavior, Equation~(\ref{ess}) is computationally attractive as it avoids matrix inversion, which is particularly advantageous for large datasets.

The behavior of ESS is most easily understood in extreme cases, namely:  
\begin{itemize}
    \item Under perfect positive correlation, \({\bf R}(\bm{\alpha}) = \bm{1}_n \bm{1}_n^\top\), leading to \(\text{ESS} = 1\). In this case, the information content of the sample is equivalent to a single observation.  
    \item Under independence, \({\bf R}(\bm{\alpha}) = {\bf I}_n\), and \(\text{ESS} = n\), indicating that each observation contributes uniquely to the information content of the dataset. 
\end{itemize}
In practice, spatial correlation typically lies between these extremes. For correlation functions satisfying \(0 \leq r(\bm{h}; \bm{\alpha}) \leq 1\), for all $\bm{h}\in\mathbb{R}^d$, one obtains \(1 \leq \text{ESS} \leq n\) \citep{acosta2018effective}. A small ESS suggests substantial redundancy due to correlation, while an ESS close to \(n\) implies that most observations provide non-overlapping information. Negative correlation may yield \(\text{ESS} > n\), although such cases are less common; see \cite{Griffith:2005} for further discussion.

Once an estimate $\widehat{\bm{\alpha}}$ has been obtained (e.g., via maximum likelihood or least squares fitting of the empirical variogram), the plug-in estimator $\widehat{\text{ESS}} = n^2 ( \bm{1}_n^\top {\bf R}(\widehat{\bm{\alpha}}) \bm{1}_n )^{-1}$ can be computed. \cite{acosta2018effective} investigated the asymptotic behavior of this estimator under a maximum likelihood framework in an increasing-domain setting, applying the delta method.

\subsection{Functional random fields}

Let \(\mathcal{H} = L^2([0,1])\) denote the Hilbert space of square-integrable functions on \([0,1]\), equipped with inner product \(\langle f, g \rangle = \int_0^1 f(t) g(t)\,\text{d}t\) and norm \(\|f\|_{\mathcal{H}} = \langle f, f \rangle^{1/2}\). A functional random field on $\mathbb{R}^d$ is a collection \(\{\chi_{{\bm{s}}} : {\bm{s}} \in \mathbb{R}^d\}\) of $\mathcal{H}$-valued random variables indexed by spatial location. Assuming \(E[\|\chi_{{\bm{s}}}\|_{\mathcal{H}}] < \infty\), the mean function \(\mu_{\bm{s}} \in \mathcal{H}\) is defined through
\[
E[\langle \chi_{\bm{s}}, f \rangle] = \langle \mu_{\bm{s}}, f \rangle, \qquad \text{for all } f \in \mathcal{H},
\]
so that \(E[\chi_{\bm{s}}(t)] = \mu_{\bm{s}}(t)\) for almost every \(t \in [0,1]\). If \(E[\|\chi_{\bm{s}}\|_{\mathcal{H}}^2] < \infty\), and   \({\bm{s}}_1, {\bm{s}}_2 \in \mathbb{R}^d\), the  covariance structure is described by the operator
\[
C_{{\bm{s}}_1,{\bm{s}}_2}(f) = E\big[ \langle \chi_{{\bm{s}}_1} - \mu_{{\bm{s}}_1}, f \rangle \, (\chi_{{\bm{s}}_2} - \mu_{{\bm{s}}_2}) \big], \qquad f \in \mathcal{H}.
\]
Pointwise, this operator admits the kernel representation
\[
C_{{\bm{s}}_1,{\bm{s}}_2}(f)(t) = \int_0^1 \sigma_{{\bm{s}}_1,{\bm{s}}_2}(t,u) f(u) \,\text{d}u,
\]
where \(\sigma_{{\bm{s}}_1,{\bm{s}}_2}(t,u) = \text{cov}[\chi_{{\bm{s}}_1}(t), \chi_{{\bm{s}}_2}(u)]\). For fixed \(t,u\), this reduces to the usual covariance of scalar random variables \citep{martinez2020recent}.

To make the statistical analysis more tractable, assumptions analogous to stationarity and isotropy in the scalar setting are often imposed.

\begin{mydef}[Weak Stationarity]
The functional random field \(\{\chi_{\bm{s}} : {\bm{s}} \in \mathbb{R}^d\}\) is weakly stationary if:
\begin{itemize}
    \item \(E[\|\chi_{\bm{s}}\|_{\mathcal{H}}^2] < \infty\) for all \({\bm{s}}\in\mathbb{R}^d\);
    \item there exists a function \(\mu \in \mathcal{H}\) such that \(\mu_{\bm{s}} = \mu\) for all \({\bm{s}}\in\mathbb{R}^d\);
    \item the covariance operator \(C_{{\bm{s}}_1,{\bm{s}}_2}\) depends on ${\bm{s}}_1, {\bm{s}}_2$ only through the difference \({\bm{s}}_1 - {\bm{s}}_2\in\mathbb{R}^d\).
\end{itemize}
\end{mydef} 

\begin{mydef}[Isotropy] 
A weakly stationary functional random field is isotropic if \(C_{{\bm{s}}_1,{\bm{s}}_2}\) depends on ${\bm{s}}_1, {\bm{s}}_2\in\mathbb{R}^d$ only through the Euclidean distance \(\|{\bm{s}}_1 - {\bm{s}}_2\|\).
\end{mydef}

These conditions are analogous to those in the scalar case: stationarity implies invariance of mean, variance, and covariance over space, while isotropy requires dependence to be the same in all directions.

A useful measure of dependence between two functional observations \(\chi_{{\bm{s}}_1}\) and \(\chi_{{\bm{s}}_2}\) is given by the trace-covariogram \citep{giraldo2011ordinary} defined as
\[
\sigma_{\text{tr}}({\bm{s}}_1,{\bm{s}}_2) = E\left[\langle \chi_{{\bm{s}}_1} - \mu_{{\bm{s}}_1}, \chi_{{\bm{s}}_2} - \mu_{{\bm{s}}_2} \rangle\right] 
= \int_0^1 \sigma_{{\bm{s}}_1,{\bm{s}}_2}(t,t) \,\text{d}t.
\]
Under isotropy, this function depends only on distance: \(\sigma_{\text{tr}}({\bm{s}}_1,{\bm{s}}_2) = \sigma_{\text{tr}}(\|{\bm{s}}_1 - {\bm{s}}_2\|)\). The corresponding kernel is denoted by \(\sigma_{\|{\bm{s}}_1 - {\bm{s}}_2\|}(\cdot,\cdot)\). Thus, for $h=\|{\bm{s}}_1 - {\bm{s}}_2\|$, we have
$$ \sigma_{\text{tr}}(h) =   \int_0^1 \sigma_{h}(t,t) \,\text{d}t. $$

A related measure of spatial variability is the trace-variogram, which quantifies the expected squared difference between two functional observations. For a pair of locations ${\bm s}_1$ and ${\bm s}_2$, the (semi) trace-variogram is defined as
$$\gamma_{\text{tr}}({\bm s}_1,{\bm s}_2) = \frac{1}{2} E\left[ \|\chi_{{\bm s}_1} - \chi_{{\bm s}_2}\|_{\mathcal{H}}^2 \right] = \frac{1}{2} \int_0^1 E\left[ \big(\chi_{{\bm s}_1}(t) - \chi_{{\bm s}_2}(t)\big)^2 \right] \, \text{d}t.$$
Under isotropy, it depends only on the distance $h = \|{\bm s}_1 - {\bm s}_2\|$, and is related to the trace-covariogram via 
\begin{equation}
    \label{eq:variog}
    \gamma_{\text{tr}}(h) = \sigma_{\text{tr}}(0) - \sigma_{\text{tr}}(h), \quad h\geq 0.
\end{equation}

\section{ESS for spatial functional data}
\label{sec:results}

\subsection{Definition and properties}

In this section, we provide a definition of the ESS for functional spatial data. We aim to adapt Definition \ref{def_ess} to the functional setting.

Suppose that a spatial functional random field is observed at $n$ locations, namely $\chi_{\bm{s}_1}, \hdots, \chi_{\bm{s}_n}$. As in the scalar case, the key question is how many independent function-valued observations would provide the same information as the current sample. While the full covariance operator provides the complete description of dependence between functional observations, the trace-covariogram offers a more parsimonious alternative by summarizing the covariance at each pair of locations into a single scalar value. We use this summary as the basis for defining the functional ESS.

\begin{mydef}
\label{def_essf}
Consider a sample $\chi_{\bm{s}_1}, \hdots, \chi_{\bm{s}_n}$ from an isotropic functional random field. The functional ESS is defined as
\begin{equation}
\label{ess_f}
\text{ESS}_{\mathcal{F}}  =  n^2 \sigma_{\text{tr}}(0)  \left(  \sum_{i=1}^n  \sum_{j=1}^n  \sigma_{\text{tr}}(\|\bm{s}_i-\bm{s}_j\|) \right)^{-1}.
\end{equation}  
\end{mydef}

The following proposition provides lower and upper bounds for $\text{ESS}_{\mathcal{F}}$. 
\begin{proposition}
\label{prop1}
Suppose that the trace-covariogram $h \mapsto \sigma_{\text{tr}}(h)$ is non-negative for all $h\geq 0$. Then, $1 \leq \text{ESS}_{\mathcal{F}}  \leq n$.
\end{proposition}
\begin{prueba}
To prove that $\text{ESS}_{\mathcal{F}}  \leq n$, we follow the same arguments used in \cite{acosta2018effective} for the scalar case. Indeed, since the trace-covariogram is non-negative,  $$  \sum_{i=1}^n  \sum_{j=1}^n \sigma_{\text{tr}}(\|\bm{s}_i-\bm{s}_j\|) =  n \sigma_{\text{tr}}(0) + 2 \sum_{i=1}^{n-1}\sum_{j=i+1}^n  \sigma_{\text{tr}}(\|\bm{s}_i-\bm{s}_j\|) \geq n\sigma_{\text{tr}}(0).$$ 
Thus, $\text{ESS}_{\mathcal{F}}  \leq n^2 \sigma_{\text{tr}}(0) (n \sigma_{\text{tr}}(0))^{-1}  =  n.$ The first part of the proof is completed.

For the second part, we note that for each fixed $t\in[0,1]$, $\{\chi_{\bm{s}}(t):  \bm{s}\in\mathbb{R}^d\}$ is an isotropic scalar random field with covariance function given by  $$\text{cov}\big[ \chi_{\bm{s}_i}(t)  ,  \chi_{\bm{s}_j}(t)\big] = \sigma_{\|\bm{s}_i - \bm{s}_j\|}(t,t).$$ It is well-known that the covariance function of an isotropic scalar random field attains its maximum value at distance zero (see, e.g., Chapter 2 in \citealp{chiles2009geostatistics}). More precisely, for all $t\in[0,1]$ and $\bm{s}_i,\bm{s}_j\in\mathbb{R}^d$, $$|\sigma_{\|\bm{s}_i - \bm{s}_j\|}(t,t)| \leq \sigma_{0}(t,t).$$ The monotony of the integral implies that $\sigma_{\text{tr}}(\|\bm{s}_i-\bm{s}_j\|) \leq \sigma_{\text{tr}}(0)$. Therefore, 
 $$\text{ESS}_{\mathcal{F}}  \geq  n^2 \sigma_{\text{tr}}(0) (n^2 \sigma_{\text{tr}}(0))^{-1}  =  1.$$
  $\qed$
\end{prueba} 

The following proposition establishes the behavior of the $\text{ESS}_{\mathcal{F}}$ in two illustrative limit cases. These are expected properties satisfied by this quantity. The proof is straightforward, and it will thus be omitted. 
\begin{proposition}
The $\text{ESS}_{\mathcal{F}}$ satisfies the following properties:
\begin{enumerate} 
\item[(1)] Suppose that $\sigma_{\|\bm{s}_i-\bm{s}_j\|}(t,t) = 0$, for all $t\in[0,1]$ and $i\neq j$. Then, $\text{ESS}_{\mathcal{F}}   = n$.
\item[(2)] Suppose that $\sigma_{\|\bm{s}_i-\bm{s}_j\|}(t,t) = \sigma_{0}(t,t)$, for all $t\in[0,1]$ and $i,j=1,\hdots,n$. Then, $\text{ESS}_{\mathcal{F}}   = 1$.
\end{enumerate}

\end{proposition}
Point (1) establishes that when the point-wise covariance between any pair of different curves is zero, the sample size should not be decreased, because all the curves contain differentiating features. Point (2) establishes that when the point-wise covariance between any pair of curves is a positive constant, i.e., even though the observations might be far apart, the covariance does not decrease (which is equivalent to perfect positive point-wise correlation between the curves),  the functional ESS is one.

The following proposition shows that adding a new independent functional observation increases the functional ESS, an intuitive and expected consequence.

\begin{proposition}
Let $\text{ESS}_{\mathcal{F}}^{(n)}$ be the functional ESS in Definition \ref{def_essf} based on $\chi_{\bm{s}_1}, \hdots, \chi_{\bm{s}_n}$. Suppose that an additional observation $\chi_{\bm{s}_{n+1}}$, independent of the existing sample, is recorded, and denote the corresponding effective sample size by $\text{ESS}_{\mathcal{F}}^{(n+1)}$. If the trace-covariogram is non-negative, then 
$$
\text{ESS}_{\mathcal{F}}^{(n+1)} > \text{ESS}_{\mathcal{F}}^{(n)}.
$$
\end{proposition}
\begin{prueba}
By independence, the trace-covariogram values involving the new functional observation satisfy
$\sigma_{\text{tr}}(\|\bm{s}_i-\bm{s}_{n+1}\|)=0$ for every $i=1,\hdots,n$. Using this fact, we can write
$$
\frac{1}{\text{ESS}_{\mathcal{F}}^{(n+1)}} 
= \frac{1}{(n+1)^2 \sigma_{\text{tr}}(0)}  
\sum_{i=1}^{n+1}  \sum_{j=1}^{n+1}  \sigma_{\text{tr}}(\|\bm{s}_i-\bm{s}_j\|)
= \frac{1}{(n+1)^2 \sigma_{\text{tr}}(0)}  
\left( \sum_{i=1}^{n}  \sum_{j=1}^{n}  \sigma_{\text{tr}}(\|\bm{s}_i-\bm{s}_j\|)  +  \sigma_{\text{tr}}(0) \right).
$$
Consequently, a straightforward calculation yields
$$
\frac{1}{\text{ESS}_{\mathcal{F}}^{(n+1)}} 
=  \frac{1}{(n+1)^2}  
\left(  \frac{n^2}{\text{ESS}_{\mathcal{F}}^{(n)}}  + 1 \right)
   \iff 
 \frac{\text{ESS}_{\mathcal{F}}^{(n)}}{\text{ESS}_{\mathcal{F}}^{(n+1)}}  =   \frac{1}{(n+1)^2}  
\left(  n^2  + \text{ESS}_{\mathcal{F}}^{(n)} \right).
$$
Finally, by Proposition \ref{prop1}, we have $\text{ESS}_{\mathcal{F}}^{(n)} \leq n < 2n+1$. It follows that $ n^2  + \text{ESS}_{\mathcal{F}}^{(n)} < n^2 + 2n +1 = (n+1)^2$. Hence,
$$
\frac{\text{ESS}_{\mathcal{F}}^{(n)}}{\text{ESS}_{\mathcal{F}}^{(n+1)}}   
<  1
\iff  
\text{ESS}_{\mathcal{F}}^{(n+1)} >  \text{ESS}_{\mathcal{F}}^{(n)}.
$$
$\qed$
\end{prueba}

\subsection{Estimation of ESS$_{\mathcal{F}}$}

The empirical trace--covariogram at distance $h$ is defined as (see, e.g., \citealp{delicado2010statistics})
\begin{equation}
\label{eq:empirical}
\widehat{\sigma}_{\text{tr}}(h)
= \frac{1}{|N(h)|}\sum_{(i,j)\in N(h)}\int_0^1   \big( \chi_{\bm s_i}(t) - \overline{\chi}(t) \big)   \big( \chi_{\bm s_j}(t) - \overline{\chi}(t) \big) \,\mathrm{d}t,
\end{equation}
where $\overline{\chi}$ denotes the sample mean function, 
$N(h)=\{(i,j): \|\boldsymbol{s}_i-\boldsymbol{s}_j\|=h\}$,
and $|N(h)|$ is the cardinality of $N(h)$. 
As in the case of scalar random fields, the distance between locations $\bm s_i$ and $\bm s_j$ does not need to match $h$ exactly; rather, all pairs satisfying $\|\boldsymbol{s}_i-\boldsymbol{s}_j\| \in (h-\delta, h+\delta)$
for some tolerance $\delta > 0$ are included in the computation.  
Since the functional observations $\chi_{\bm s_i}$ are available only at a finite set of points $t_{i1}, \ldots, t_{im}$, the evaluation of \eqref{eq:empirical} requires a numerical approximation of the integral.

In practice, the estimator \eqref{eq:empirical} is evaluated at a finite set of lag distances $h_1, \ldots, h_q$. A parametric model can then be fitted to the empirical estimates $\widehat{\sigma}_{\mathrm{tr}}(h_1), \ldots, \widehat{\sigma}_{\mathrm{tr}}(h_q)$, for example via a least squares approach, yielding $\widehat{\sigma}_{\mathrm{tr}}(h)$ for all $h \geq 0$. Consequently, a plug-in estimator of $\mathrm{ESS}_{\mathcal{F}}$ in \eqref{ess_f} is defined through
\begin{equation}\label{eq_ess_est}
\widehat{\text{ESS}}_{\mathcal{F}}= n^2 \widehat{\sigma}_{\text{tr}}(0)  \left(  \sum_{i=1}^n  \sum_{j=1}^n  \widehat{\sigma}_{\text{tr}}(\|\bm{s}_i-\bm{s}_j\|) \right)^{-1}.
\end{equation}

In Section \ref{sec:data}, we apply the estimation procedure described above using the \texttt{R} package \texttt{fdagstat} \citep{fdagstat}. Specifically, we estimate the trace-variogram, which, through the relationship in Equation (\ref{eq:variog}), allows us to obtain an estimate of the  trace-covariogram, and consequently, of the functional ESS.

\subsection{Example: the FAR(1) model}

To provide insight into the proposed definition, we consider a functional time series example, namely a random field indexed by a subset of $\mathbb{R}$ and governed by an autoregressive dependence structure. While the paper primarily addresses spatially indexed functional data, this temporally indexed example enables closed-form derivations of $\mathrm{ESS}_{\mathcal{F}}$ and supports a more detailed illustration of its properties, mirroring ESS results for autoregressive processes in the scalar setting \citep{cressie2015statistics}.

A functional autoregressive process of order one (FAR(1)) is given by
$$
\chi_n = \Psi(\chi_{n-1}) + \epsilon_n,  \quad n\in\mathbb{N},
$$
where $\Psi: \mathcal{H}\rightarrow \mathcal{H}$ is a bounded linear operator and $\{\epsilon_n: n\in\mathbb{N}\}$ is a zero-mean white noise process in $\mathcal{H}$ \citep{bosq2000linear}.  In particular, let us consider $\Psi$ of the form
$\Psi(\chi)(t) = \int_0^1 \psi(t,s)\chi(s)\,ds$, where $\psi$ is square-integrable, continuous, symmetric, and positive, so that $\Psi$ is a compact, self-adjoint, positive operator.
By Mercer’s theorem, $\psi$ admits the spectral decomposition
$$
\psi(t,s) = \sum_{k = 1}^\infty \lambda_k \varphi_k(t)\varphi_k(s), \quad t,s\in[0,1],
$$
where $\lambda_1,\lambda_2,\hdots$ are (non-negative) eigenvalues and $\varphi_1,\varphi_2,\hdots$ are orthonormal eigenfunctions. For stationarity, it is additionally required that
$\|\Psi\| = \sup_k \lambda_k < 1$ (the norm of the operator must lie within the unit disk). Observe that, for simplicity, the process is taken to have zero mean.

The 1-lag trace-covariogram is 
$
\sigma_{\text{tr}}(1) = E\big[\langle \chi_n, \chi_{n-1} \rangle\big]     
=  E\big[ \langle \Psi(\chi_{n-1}), \chi_{n-1} \rangle \big] 
+  E\big[\langle \epsilon_n, \chi_{n-1} \rangle].
$
The second term is zero due to the independence between $\chi_{n-1}$ and $\epsilon_n$. To characterize the first term, notice that
expanding $\Psi$ via its Mercer representation yields
$$
\langle \Psi(\chi_{n-1}), \chi_{n-1} \rangle
= \sum_{k=1}^\infty \lambda_k \, \langle \chi_{n-1}, \varphi_k \rangle^2.
$$
Under stationarity, each scalar projection $\langle \chi_n, \varphi_k \rangle$, $k\in\mathbb{N}$,
is a process following an AR(1) structure with autoregressive coefficient $\lambda_k$, and hence its variance is given by the standard AR(1) formula
$$
E\big[\langle \chi_{n-1}, \varphi_k \rangle^2\big]
= \frac{\eta_k^2}{1 - \lambda_k^2},
$$
where $\eta_k^2$ denotes the variance of the noise component along $\varphi_k$, i.e., $\eta_k^2=E\left[\langle \epsilon_n,\varphi_k \rangle^2 \right]$.
Consequently,
$$
\sigma_{\text{tr}}(1) = \sum_{k=1}^\infty \lambda_k \, \frac{\eta_k^2}{1 - \lambda_k^2}.
$$
In general, the $h$-lag trace-covariogram, $\sigma_{\text{tr}}(h) = E\big[\langle \chi_n, \chi_{n-h} \rangle\big]$, can be obtained similarly. Indeed, 
iterating the AR(1) recursion along each eigenfunction implies that, for every $k\in\mathbb{N}$,
$$
E\big[\langle \chi_n, \varphi_k \rangle
\langle \chi_{n-h}, \varphi_k \rangle\big]
= \lambda_k^h \,
E\big[\langle \chi_n, \varphi_k \rangle^2\big].
$$
Therefore,
$$
\sigma_{\text{tr}}(h) = \sum_{k=1}^\infty \lambda_k^h \, \frac{\eta_k^2}{1 - \lambda_k^2},
$$
showing that the $h$-lag trace-covariogram decays geometrically with $h$,
at rates determined by the eigenvalues $\lambda_k$ of the kernel.

Building on the previous calculations, the functional ESS for a sample $\chi_1, \dots, \chi_n$ from the FAR(1) model introduced above is given by
$$ \text{ESS}_{\mathcal{F}}  =       
\frac{  n^2 \sum_{k=1}^\infty \frac{\eta_k^2}{1-\lambda_k^2}      }{            \sum_{k=1}^\infty \frac{\eta_k^2}{1-\lambda_k^2}  \left(  \sum_{i=1}^n  \sum_{j=1}^n   \lambda_k^{  |i-j| } \right)     }.
$$
To rewrite this expression in a more interpretable form, introduce, for each $k\in\mathbb{N}$, the normalized weights
$$p_k = \frac{\eta_k^2}{(1-\lambda_k^2)} \left( \sum_{r=1}^\infty \frac{\eta_r^2}{(1-\lambda_r^2)} \right)^{-1},$$
and the marginal effective sample sizes given by
$$\text{ESS}_k = n^2 \left(  \sum_{i=1}^n  \sum_{j=1}^n   \lambda_k^{  |i-j| } \right)^{-1}.$$
Here, the term \emph{marginal} indicates that this quantity corresponds to the ESS of the scalar process
$\langle \chi_n, \varphi_k \rangle$.
Therefore,
\begin{equation}
    \label{harmonic_mean}
    \text{ESS}_{\mathcal{F}}  =  \left( \sum_{k=1}^\infty  \frac{ p_k}{\text{ESS}_k}\right)^{-1},
\end{equation}
which corresponds to a weighted harmonic mean of the marginal effective sample sizes.

\subsection{Numerical study for the FAR(1) model}

\begin{figure}
    \centering
    \includegraphics[scale=0.08]{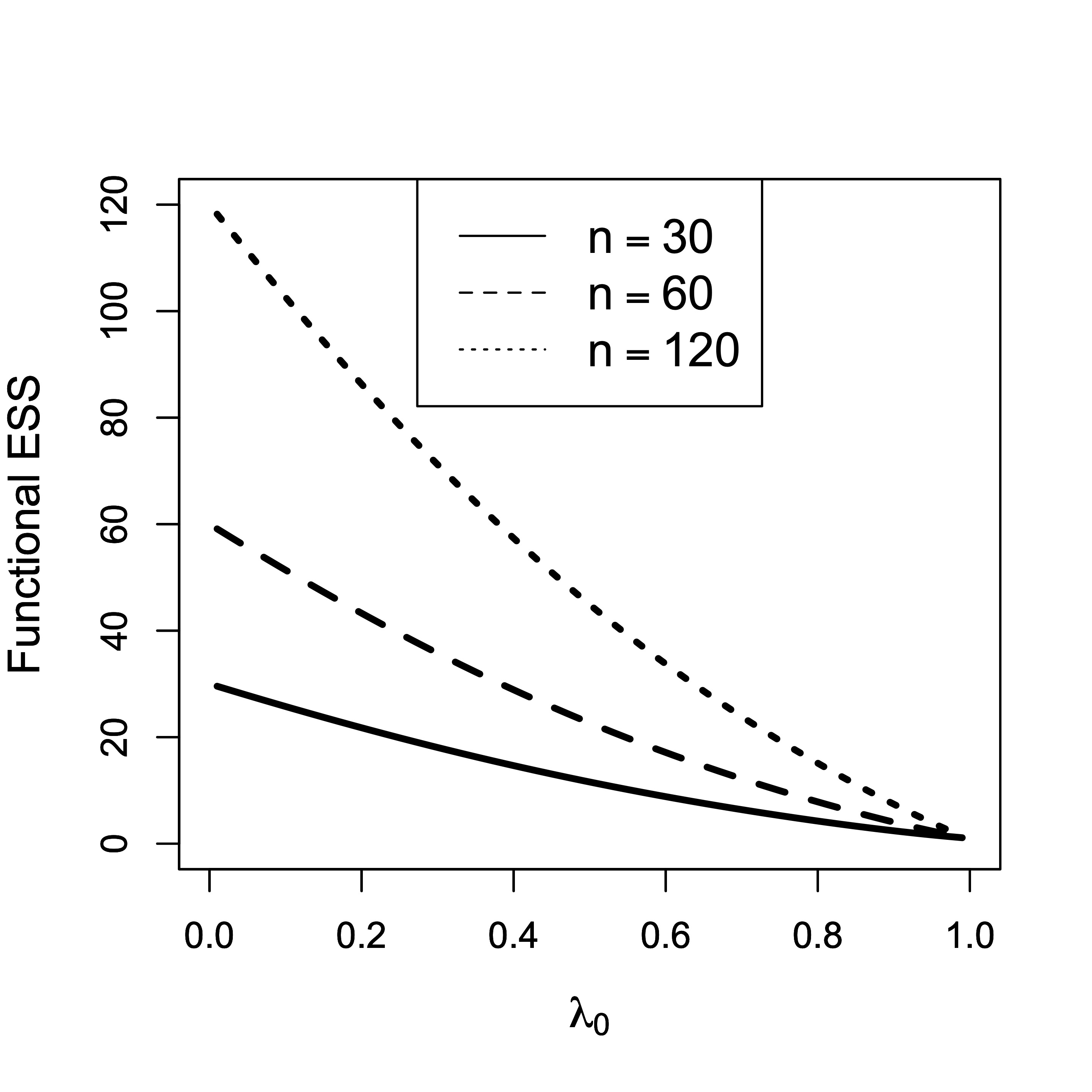}  \includegraphics[scale=0.08]{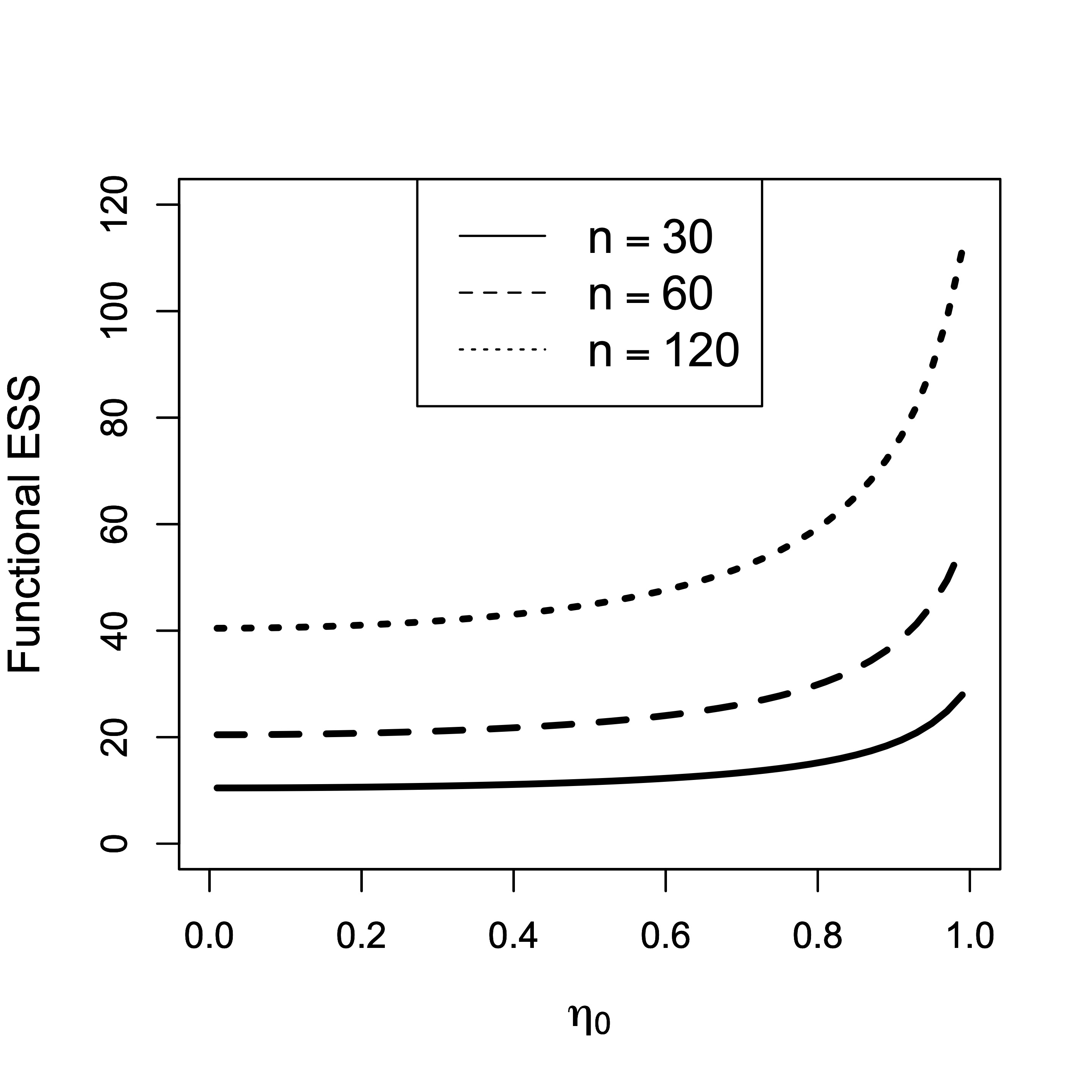}
    \caption{Functional ESS under the FAR(1) model. Left: dependence on $\lambda_0 \in (0,1)$ with $\lambda_k = \lambda_0^k$ and fixed $\eta_k = (1/2)^k$. Right: dependence on $\eta_0 \in (0,1)$ with $\eta_k = \eta_0^k$ and fixed $\lambda_k = (1/2)^k$. Results are shown for $n = 30, 60,$ and $120$.}
    \label{fig:far}
\end{figure}

Figure \ref{fig:far} illustrates a numerical study of the functional ESS under the FAR(1) model. In the left panel, the functional ESS is displayed as a function of $\lambda_0 \in (0,1)$, where the eigenvalues of the autoregressive operator are specified as $\lambda_k = \lambda_0^k$. This parametrization controls the overall strength of temporal dependence, with larger values of $\lambda_0$ corresponding to slower decay of dependence across the eigen-directions of the operator. The functional ESS is computed for three sample sizes, $n = 30, 60,$ and $120$. Throughout this analysis, the noise variances are fixed according to $\eta_k = (1/2)^k$.

The decreasing behavior of the functional ESS as a function of $\lambda_0$ is fully consistent with the underlying structure of the FAR(1) model. Larger values of $\lambda_0$ induce stronger temporal autocorrelation along all eigen-directions. As a consequence, successive functional observations become increasingly redundant, leading to a reduction in the effective amount of information. Conversely, as $\lambda_0 \to 0$, the dependence structure weakens uniformly across eigen-directions and the process approaches a functional white-noise regime. In this case, temporal dependence becomes negligible and the functional ESS approaches the nominal sample size.

The right panel of Figure \ref{fig:far} presents a complementary analysis in which the noise structure of the FAR(1) model is varied while the dependence structure is held fixed. Specifically, the noise variances are defined as $\eta_k = \eta_0^k$, with $\eta_0 \in (0,1)$, and the autoregressive coefficients are fixed as $\lambda_k = (1/2)^k$. As in the left panel, results are shown for $n = 30, 60,$ and $120$.

A different mechanism explains the increase of the functional ESS as $\eta_0$ approaches one. Larger values of $\eta_0$ imply a slower decay of the noise variances $\eta_k^2$, which results in a larger number of terms contributing non-negligibly to the weighted harmonic mean in Equation (\ref{harmonic_mean}). In this setting, the functional ESS reflects the aggregation of a broader collection of marginal effective sample sizes $\text{ESS}_k$, many of which correspond to higher-order terms with smaller autoregressive coefficients $\lambda_k$ and therefore weaker temporal dependence. As a result, these larger marginal effective sample sizes receive greater weight in the harmonic mean, leading to an overall increase in the functional ESS.

\section{Real data illustration}
\label{sec:data}
 
We analyze geometric vertical velocity data from the NCEP Global Ocean Data Assimilation System (GODAS), available at \url{https://psl.noaa.gov}. This variable, measured in m$\cdot$s$^{-1}$, represents the rate of vertical movement of ocean water parcels relative to the Earth’s surface (see \citealp{behringer1998improved}), and is relevant for understanding vertical transport and mixing in the ocean.

Our analysis focuses on the monthly mean for January 2024 and on a region of the Pacific Ocean bounded by latitudes 35$^\circ$N–45$^\circ$N and longitudes 135$^\circ$W–155$^\circ$W. This localized spatial window facilitates the characterization of the data through an isotropic random field assumption.  The dataset has a spatial resolution of $0.333^\circ$ in latitude and $1.0^\circ$ in longitude.

\begin{figure}
    \centering
    \includegraphics[scale=0.075]{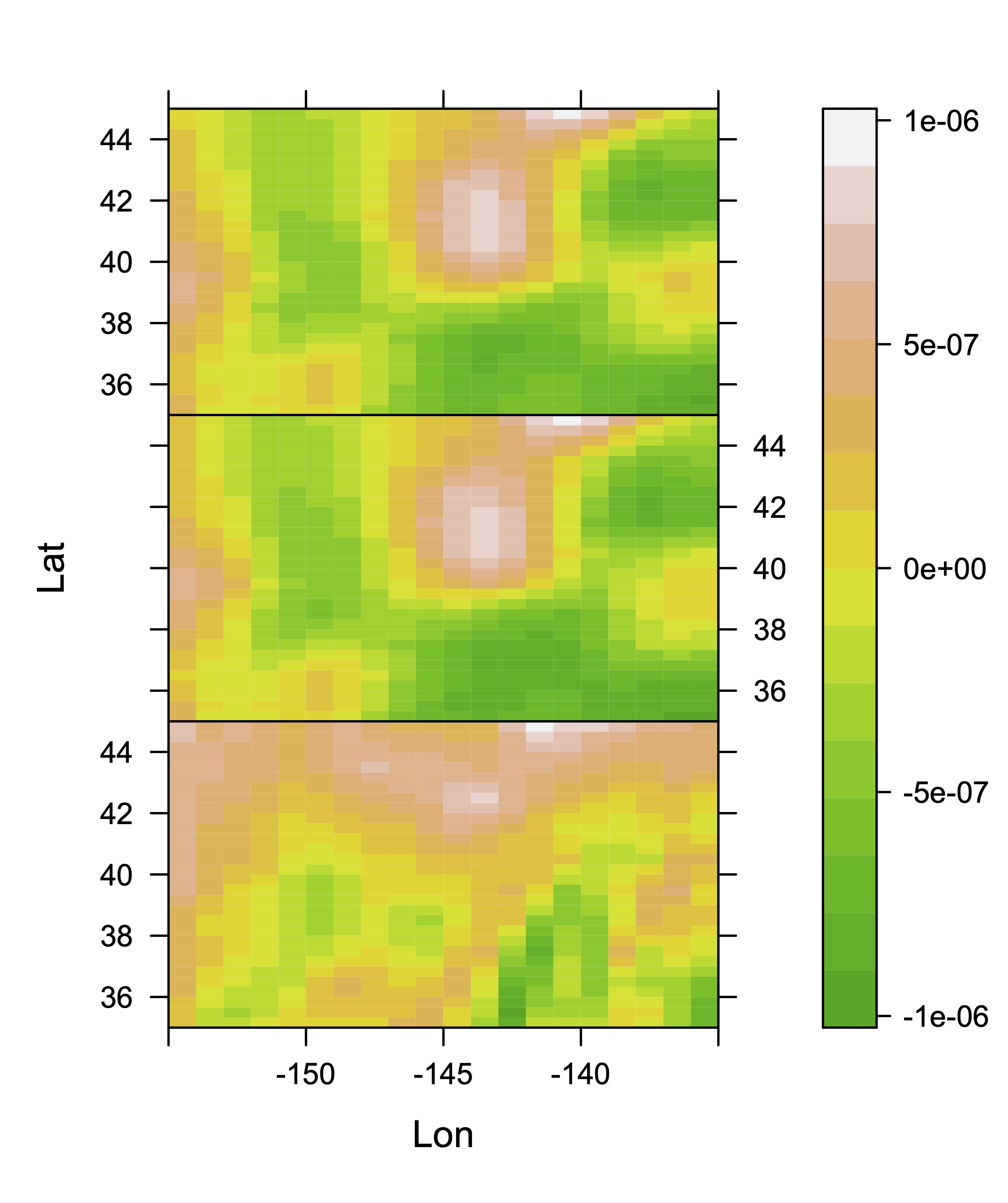}  \includegraphics[scale=0.085]{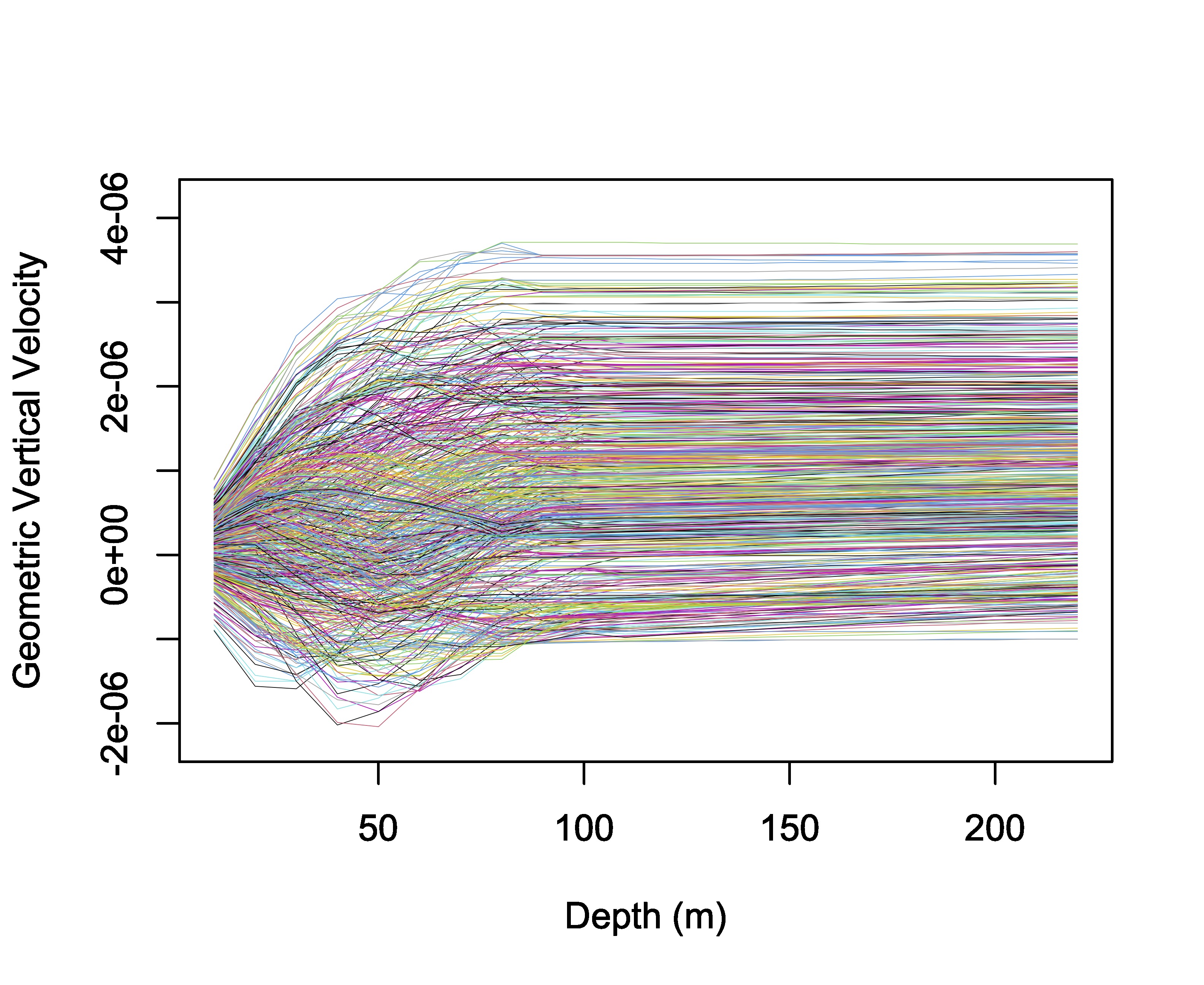}
    \caption{Geometric vertical velocity (in m$\cdot$s$^{-1}$) over a region of the Pacific Ocean for January 2024. Left: spatial maps at three depth levels (20 m, 10 m, and 1 m, from top to bottom). Right: vertical profiles of velocity as a function of depth (in m) for each of the 600 spatial locations.}
    \label{fig:heatmaps}
\end{figure}

The resulting dataset consists of 600 spatial locations. At each location, a geometric vertical velocity is considered at 22 equispaced vertical levels, ranging from 10 m to 220 m in 10 m increments. Therefore, it is treated as a function of the level, giving rise to the functional nature of the dataset. Although the curves in this dataset are defined on a different domain than the interval $[0,1]$ used in Section \ref{sec:background}, the same functional formulation applies. Figure \ref{fig:heatmaps} (left panel) illustrates the variable of interest at three representative levels, while Figure \ref{fig:heatmaps} (right panel) displays the full set of 600 functional curves of geometric vertical velocity. Together, these visualizations provide insight into the spatial and vertical structure of the velocities.

For our analysis, the geographical coordinates are first projected onto the plane using a sinusoidal projection. This allows us to work within the framework of a functional random field indexed by a planar region. The next step is to fit a covariance model to the data, for which we consider the following parametric structures for the trace-covariogram:
\begin{itemize}
 \item Exponential: $\sigma_{\text{tr}}(h) = \sigma^2 \exp(-h/\alpha)$.
 
    \item Spherical:  \begin{equation*}
\sigma_{\text{tr}}(h)  = 
\begin{cases} 
\sigma^2 \left[ 1 - \frac{3}{2} \left( h / \alpha \right) + \frac{1}{2} \left( h / \alpha \right)^3 \right] & \text{ if } 0 \le h \le \alpha \\
0 & \text{ otherwise } 
\end{cases}.
\end{equation*}
   
    \item Gaussian:  $\sigma_{\text{tr}}(h) = \sigma^2 \exp(-h^2/\alpha^2)$.
\end{itemize}

\begin{figure}
    \centering
    \includegraphics[scale=0.09]{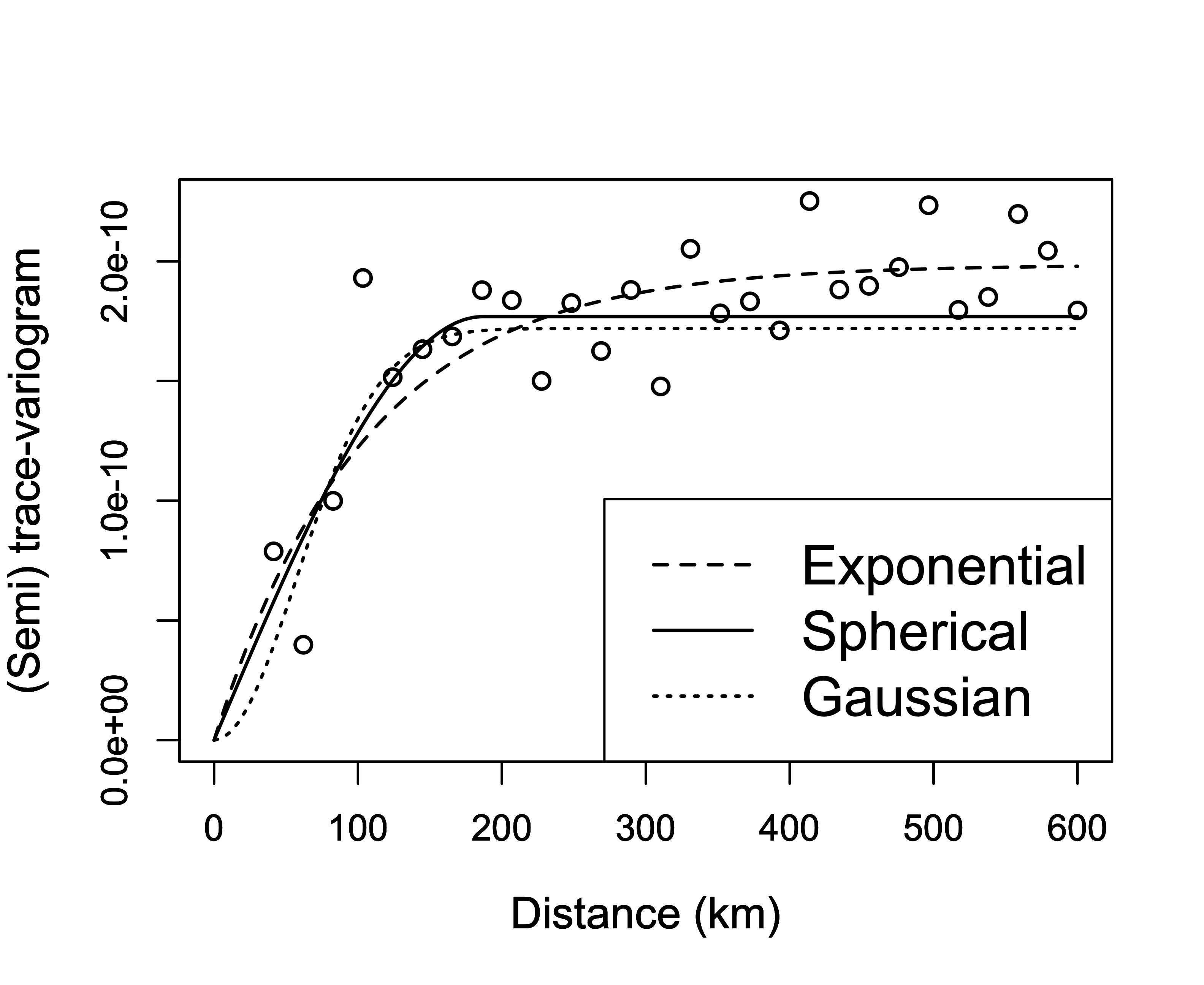}
    \caption{Empirical (circles) and fitted (curves) (semi) trace-variograms for the geometric vertical velocity dataset.  }
    \label{fig:variogram}
\end{figure}

Parameters are estimated using a least squares approach by fitting the aforementioned theoretical variogram models that minimize the quadratic discrepancy with respect to the empirical trace-variogram. The resulting parameter estimates are $(\widehat{\sigma}^2,\widehat{\alpha}) = (1.985\times 10^{-10}, 104.4)$ for the Exponential model, $(\widehat{\sigma}^2,\widehat{\alpha}) = (1.769\times 10^{-10}, 186.8)$ for the Spherical model, and $(\widehat{\sigma}^2,\widehat{\alpha}) = (1.719\times 10^{-10}, 81.12)$ for the Gaussian model. Figure \ref{fig:variogram} shows the empirical and fitted variograms, indicating a satisfactory agreement between observed and theoretical quantities. We also considered models including a nugget effect; however, it was estimated as zero in all experiments and is therefore not reported.

The estimated functional ESS values for the Exponential, Gaussian and Spherical models are $41.92$, $105.4$ and $102.2$, respectively. For both the Spherical and Gaussian models, the results suggest that approximately 17\% of the original sample suffices to capture the essential statistical information. In contrast, the Exponential model indicates a more substantial reduction in effective sample size. 

\begin{figure}
    \centering
    \includegraphics[scale=0.1]{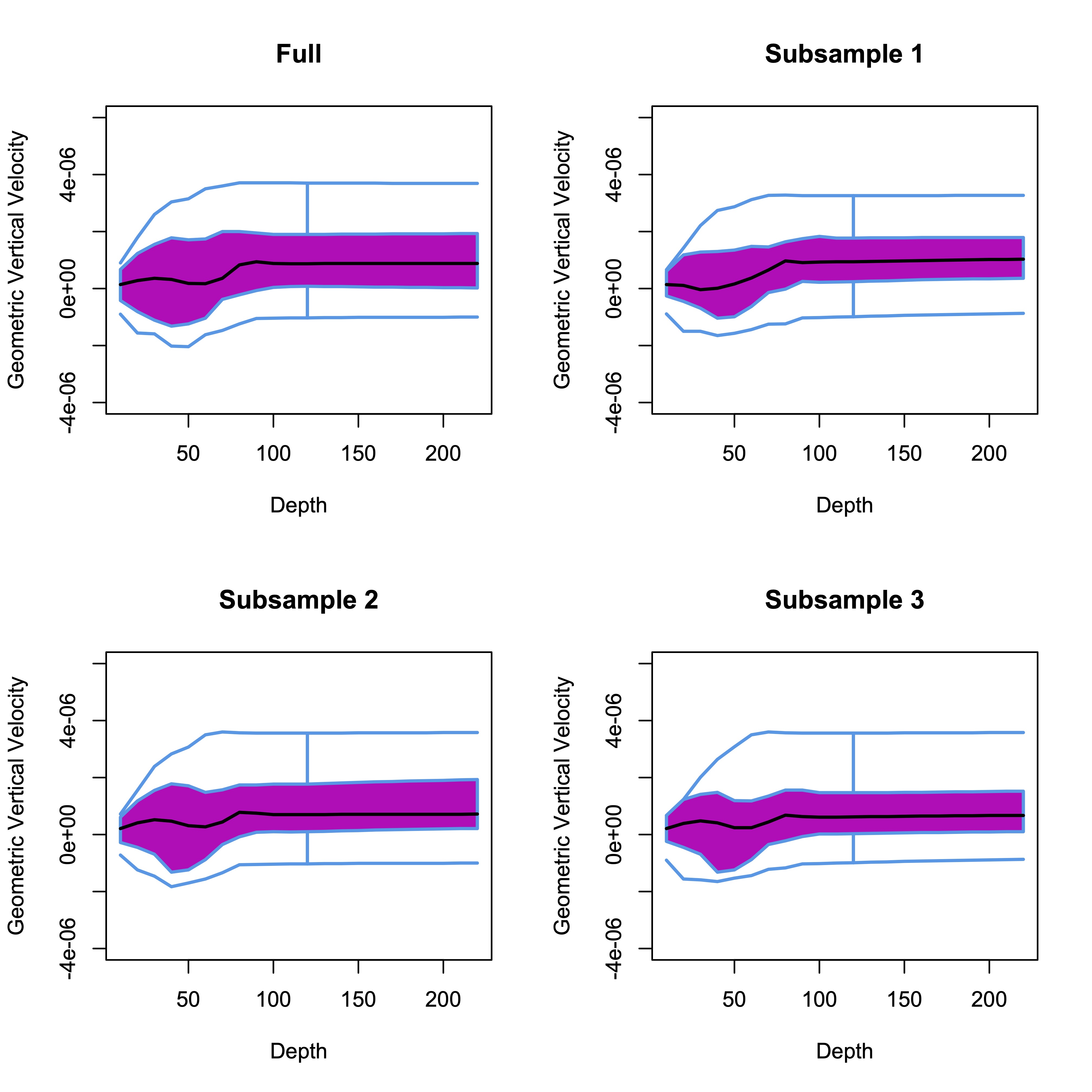}
    \caption{Functional boxplots of the full dataset and of three independent subsamples of size 106, corresponding to the maximum functional ESS value obtained across all models.}
    \label{fig:fboxplots}
\end{figure}

To illustrate these findings, we construct a functional boxplot (Figure \ref{fig:fboxplots}) using the full dataset and compare it with boxplots obtained from three randomly selected subsamples of size 106, corresponding to the maximum ESS value obtained across all models. The resulting plots show only minor differences, providing visual support for our conclusions. The largest variations are observed at lower depths, where the functional boxplot of the full dataset exhibits the most pronounced changes. For example, the subsamples capture the envelope of the 50\% central region accurately. Similar observations apply to other features of the plot, in particular, the median curve and the maximum non-outlying envelope are also well represented.

To quantify the similarity between the full sample and the subsamples, and to complement the visual assessment provided by functional boxplots, five numerical measures derived from the structure of the functional boxplot were considered. First, the stability of the functional center was examined using two distance measures between the functional medians of the full sample and each subsample: the mean squared distance between medians (MD-L2), which summarizes overall discrepancies across the domain, and the maximum discrepancy between medians (MD-Sup), which captures the largest pointwise deviation. Second, the stability of functional dispersion was assessed by comparing the 50\% central regions using two analogous measures: the mean discrepancy in the width of the central region (CRD-Mean), which reflects average changes in functional variability, and the maximum discrepancy in the width of the central region (CRD-Sup), which identifies localized differences in dispersion. Finally, the representativeness of each subsample was evaluated through the central inclusion proportion (CIP), defined as the proportion of curves in the subsample that lie within the 50\% central region of the full sample. Taken together, these five measures capture complementary aspects of functional location, dispersion, and representativeness, and offer a quantitative evaluation of the stability of the functional structure under subsampling, as summarized in Table \ref{tab:summary}.

In addition to the three illustrative subsamples, we conduct a more comprehensive resampling experiment. Specifically, we randomly draw 1,000 independent subsamples of size 106 from the full dataset and compute the same five indicators for each subsample. We then report the average value of each indicator across the 1,000 subsamples (see the last row in Table \ref{tab:summary}), offering a systematic summary that complements the analysis.

\begin{table}[]
    \centering
    \renewcommand{\arraystretch}{1.2} 
    \setlength{\tabcolsep}{8pt}  
    \begin{tabular}{lccccc} \hline \hline 
    & MD-L2  & MD-Sup  & CRD-Mean & CRD-Sup & CIP \\ \hline 
Subsample 1 & $2.30\times 10^{-7}$ &  $2.80\times 10^{-7}$  &  $2.81\times 10^{-7}$  &  $4.90\times 10^{-7}$  &  0.518 \\

Subsample 2 & $4.04\times 10^{-7}$ & $5.40\times 10^{-7}$ & $1.48\times 10^{-7}$ & $5.30\times 10^{-7}$ & 0.528\\

Subsample 3 & $1.51\times 10^{-7}$ & $2.70\times 10^{-7}$ & $2.38\times 10^{-7}$ & $4.30\times 10^{-7}$ & 0.575 \\   \hline 
{\small Avg. (1,000 subsamples)} & $2.37\times 10^{-7}$  & $4.16\times 10^{-7}$ & $2.56 \times 10^{-7}$ &
  $4.69 \times 10^{-7}$  &   0.513 \\
\hline \hline 
    \end{tabular}
    \caption{Numerical comparison between the full sample, three illustrative subsamples, and the average over 1,000 randomly drawn subsamples, based on five functional boxplot–based measures. MD-L2 and MD-Sup quantify global and maximum discrepancies between functional medians, CRD-Mean and CRD-Sup measure average and maximum differences in the width of the 50\% central regions, and CIP denotes the proportion of subsample curves contained within the 50\% central region of the full sample.}
    \label{tab:summary}
\end{table}

\begin{figure}
    \centering
    \includegraphics[scale=0.08]{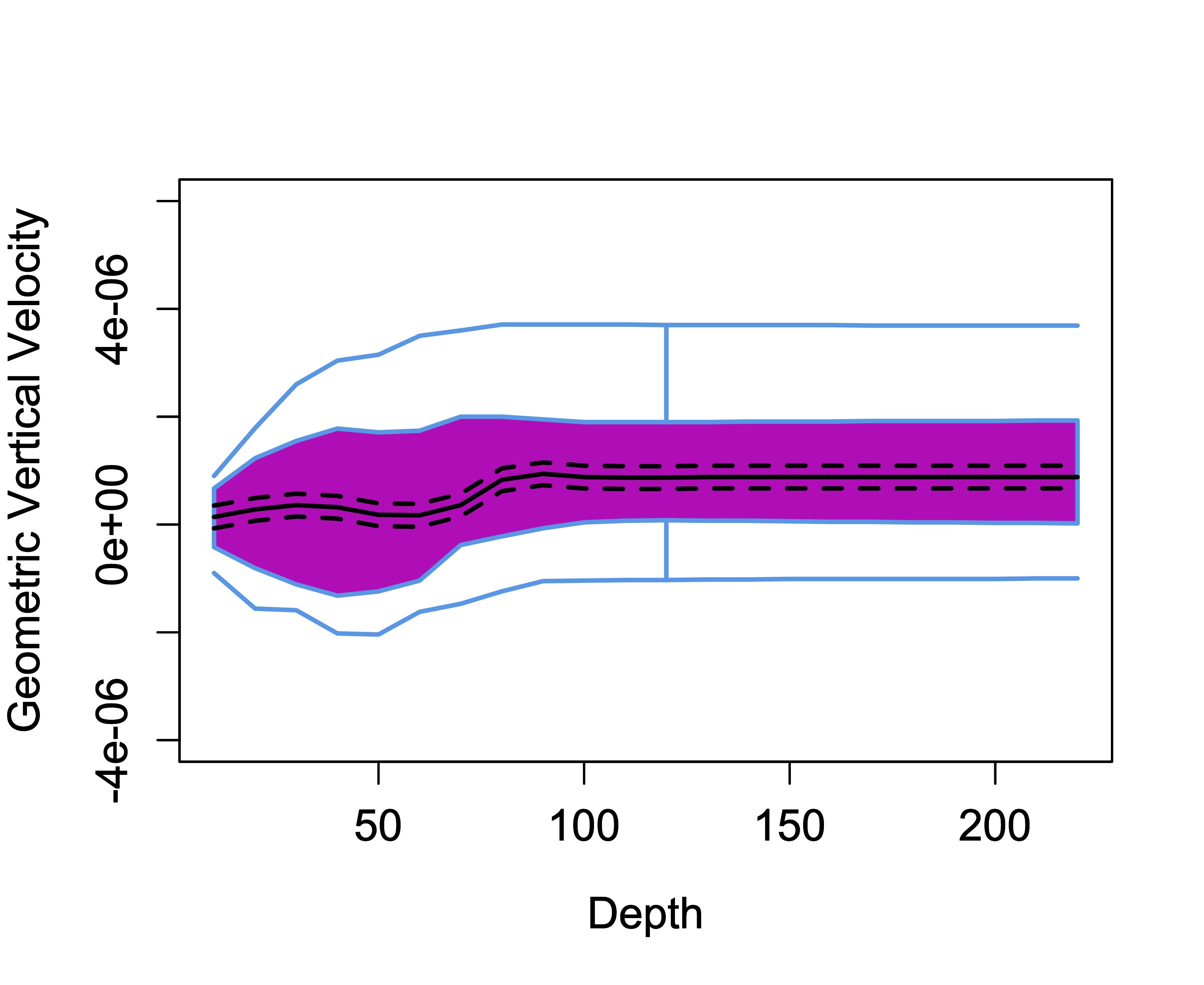}
    \caption{Functional boxplot of the full sample, with an additional dashed band around the functional median. The band extends by $\pm 2.11\times 10^{-7}$, corresponding to the mean absolute discrepancy between the functional median of the full sample and the medians computed across 1,000 subsamples.}
    \label{fig:median_error}
\end{figure}

Notice that the CIP values are close to 0.5, as expected. Moreover, although the discrepancy measures have small numerical values, their magnitude is not directly interpretable without an appropriate scale. To provide a visual reference, Figure \ref{fig:median_error} shows the functional boxplot of the full sample, with a dashed band added around its median. The band extends by ${\pm 2.11\times 10^{-7}}$, which represents the mean absolute discrepancy between the functional medians, averaged over the 1,000 subsamples. This graphical representation demonstrates that the observed discrepancy between the medians of the full sample and the subsamples is small relative to the overall variability of the functional data, thereby reinforcing the conclusion that the subsamples closely resemble the full sample. The discrepancies in the widths of the central regions support the same conclusion.

\section{Concluding remarks}
\label{sec:conclusions}

We proposed an approach to extend the notion of ESS from the purely scalar random field setting to functional random fields. Our definition of functional ESS, based on the trace-covariogram as a measure of spatial association, preserves the intuitive properties of its scalar counterpart, providing an interpretable and meaningful measure.

The results from our real-data illustration on geometric vertical velocities may be particularly relevant for studies of ocean circulation.  Functional ESS can guide the selection of a reduced set of observations, thereby decreasing storage requirements or measurement effort in practical settings where data collection is costly or sparse, while still preserving essential statistical information. In our study, for example, selecting just 17\% of the full sample retains the key statistical characteristics of the complete dataset. 

Representing vertical profiles as functional observations allows the approach to accommodate non-stationarities along ocean depth. In our experiments, such non-stationarities manifest as stronger velocity fluctuations near the surface and greater stability at depth, effectively represented by the functional framework.

A natural extension of the results presented in this paper is to consider multivariate spatial functional random fields, where the functional ESS would need to account not only for the autocorrelation of each individual component of the random field but also for the cross-correlation between fields. Similarly, the development of a functional ESS for anisotropic random fields, in which spatial dependence varies across directions, is an important topic for future research. Another promising direction concerns the uncertainty quantification of the sample functional ESS. This could be achieved by characterizing its asymptotic properties, which are contingent upon the limiting behavior of the empirical trace-covariogram. To the best of our knowledge, this remains an open problem that warrants further study.

\section*{Funding}

This work was funded and supported by the National Agency for Research and Development of Chile through grants ANID Fondecyt 1251154 (A. Alegr\'ia), ANID Fondecyt 1230012 (R. Vallejos) and ANID CIA250006 (R. Vallejos). J. Mateu is partially funded by grant PID2022-141555OB-I00 from Ministry of Science and Innovation. 

\section*{Conflicts of Interest}

The authors declare no conflicts of interest.

\section*{Data Availability Statement}

The data supporting the findings of this study, obtained from the NCEP Global Ocean Data Assimilation System (GODAS), are openly available at \url{https://psl.noaa.gov}.

\bibliography{sample}
\bibliographystyle{apalike}

\end{document}